\documentclass[aps,pre,reprint,showpacs,showkeys]{revtex4-1}
\usepackage{hyperref}
\usepackage{graphicx}

\begin{document}

\title{Externally-driven transmission and collisions of domain walls in ferromagnetic wires}
\author{Andrzej Janutka}
\email{Andrzej.Janutka@pwr.wroc.pl}
\affiliation{Institute of Physics, Wroclaw University of Technology, Wybrze\.ze Wyspia\'nskiego 27, 50-370 Wroc{\l}aw, Poland}

\begin{abstract}
Analytical multi-domain solutions to the dynamical (Landau-Lifshitz-Gilbert) equation 
 of a one-dimensional ferromagnet including an external magnetic field and spin-polarized electric 
 current are found using the Hirota bilinearization method. A standard approach to solve the Landau-Lifshitz 
 equation (without the Gilbert term) is modified in order to treat the dissipative dynamics.
 I establish the relations between the spin interaction parameters (the constants of exchange, 
 anisotropy, dissipation, external-field intensity, and electric-current intensity) and the domain-wall parameters
 (width and velocity) and compare them to the results of the Walker approximation and micromagnetic simulations.
 The domain-wall motion driven by a longitudinal external field is analyzed with especial relevance to
 the field-induced collision of two domain walls. I determine the result of such a collision 
 (which is found to be the elastic one) on the domain-wall parameters below and above the Walker breakdown
 (in weak- and strong-field regimes). Single-domain-wall dynamics in the presence of an external 
 transverse field is studied with relevance to the challenge of increasing the domain-wall velocity
 below the breakdown.  
\end{abstract}


\maketitle
\newpage

\section{Introduction}

Description of the magnetic-field- and electric-current-induced motions of domain walls (DWs)  
 in nanowires has became a hot topic because of novel methods of storing and switching the (magnetically
 encoded) binary information. These proposals offer a progress in the miniaturization of memory and logic elements,
 utilizing crucial advantages of magnetic information encoding, (when a bit is identified with a single
 magnetic domain). Such information is insensitive to the voltage fluctuations while its maintenance 
 does not cost any energy, which enables the data processing with the production of a small heat amount.
 Currently investigated random-access memories are built of metallic nanowires, formed into a parallel-column structure, 
 which store magnetic domains separated by DWs. Such a three-dimensional (3D) magnetic system 
 has the potential of storing more information than devices based on 2D systems, like 
 hard-disk drives or electronic memories, in a given volume \cite{par08,hay08}. Also,
 an interesting concept of logical operation via transmitting magnetic DWs through nanowires
 of specific geometries is being developed \cite{all02,all07}. I mention that ferroelectric nanosystems 
 offer similar capabilities while their basic properties are studied with the same dynamical 
 (Landau-Lifshitz-Gilbert) equation even though the effects of electroelastic coupling are strong \cite{sco07,gru08}. 

In order to write and switch information, one can move the DWs via the application 
 of an external magnetic field (parallel to the easy-axis)
 or via the application of a voltage which induces the spin-polarized electric current 
 through the DW. The directions of the field-driven motion are different for the tail-to-tail and head-to-head
 DWs; thus the magnetic field induces the DW collisions, while the direction of the voltage-driven motion 
 uniquely corresponds to the current direction. Field-driven motion and current-driven (below
 the Walker breakdown) motion are possible due to the magnetic dissipation and its description demands
 inclusion of the Gilbert term into the Landau-Lifshitz (LL) equation.
 However, existing many-domain analytical solutions to the LL equation 
 do not include the dissipation \cite{bog80,hir82}, while approximate solutions using the Walker ansatz 
 describe a single DW only \cite{sch74}. Since the parameters of the DW solutions to the Landau-Lifshitz-Gilbert 
 (LLG) equation determine accessible values of technological characteristics (e.g. the minimal domain length, 
 the bit-switching time, etc.), it is of interest to know the analytic DW solution to the LLG equation.
 Knowledge of the many-domain solution is of importance for preventing unwanted DW collisions which can result
 in an instability of the record and it enables verification of DW-collision simulations regardless
 of the internal structure or the geometry of the simulated system. 

In the present paper, I perform an analytical study of the dynamics of multi-domain systems including 
 the dissipation. The dynamical LLG system is bilinearized following the Hirota method of solving nonlinear equations 
 and it is extended, via doubling the number of freedom degrees and the number of equations,
 into a time-reversal invariant form. The field solving the extended system contains proper 
 and virtual (unphysical) dynamical variables and the physical components
 of the solution are shown to satisfy the pimary LLG system in a relevant time regime.
 In particular, aiming to analytically describe the field-induced collision, 
 I establish asymptotic three-domain magnetization profiles (relevant in the time limits $t\to\pm\infty$). 
 With connection to the phenomenon of the Walker breakdown, (a cusp in the dependence of
 the DW velocity on the external field and in its dependence on the current intensity), 
 I modify the LLG model in a way to make it applicable below the breakdown.
 I reduce it into a model of plane rotators. In this weak-field regime, the spin alignment in the DW area 
 is saturated to a plane while, above the breakdown (the strong-field regime), the spins rotate about the
 magnetic-field axis (the easy axis). These considerations supplement micromagnetic simulations 
 of the DW collisions in terms of the studied DW-parameter regimes \cite{kun09}. Within the present method, 
 I verify the Walker-ansatz predictions on the current-driven DW motion above and below the breakdown, 
 (including adiabatic and non-adiabatic parts of the spin-transfer torque) \cite{zha04}-\cite{vol08}, thus,
 showing the applicability of the present formalism to the field-driven motion of multi-domain systems with 
 an additional voltage applied. With relevance to the challenge of controlling the maximum DW velocity
 below the Walker breakdown, we analyze the longitudinal-field-driven motion and current-driven motion
 of the DWs in the presence of an additional perpendicular (with respect to the easy axis) field. 
 
A complementary study of the DW collision of magnetic DWs in the subcritical regime is performed in a separate
 paper \cite{jan10}.
   
In Sec. II, I extend the dissipative equations of motion of the ferromagnet 
 in such a way as to make equations applicable to the unlimited range of time $t\in(-\infty,\infty)$. 
 Section III is devoted to the analysis of its single- and double-DW solutions in the presence 
 of a magnetic field and an electric current. I study the field-induced collision in detail. 
 The plane-rotator approach to the DW dynamics is described in Sec. IV. In Sec. V,  
 consequences of the application of an external field perpendicular to the easy axis
 for the DW statics and dynamics are considered. Conclusions are given in Sec. VI.

\section{Dynamical equations}
 
The dynamics of the magnetization vector ${\bf m}$ ($|{\bf m}|=M$) in the 1D ferromagnet 
 is described with the LLG equation
\begin{eqnarray}
\frac{\partial{\bf m}}{\partial t}=\frac{J}{M}{\bf m}\times\frac{\partial^{2}{\bf m}}{\partial x^{2}}
+\gamma{\bf m}\times{\bf H}+\frac{\beta_{1}}{M}({\bf m}\cdot\hat{i}){\bf m}\times\hat{i}
\nonumber\\
-\frac{\beta_{2}}{M}({\bf m}\cdot\hat{j}){\bf m}\times\hat{j}
-\delta\frac{\partial{\bf m}}{\partial x}
-\frac{\delta\beta}{M}{\bf m}\times\frac{\partial{\bf m}}{\partial x}
\nonumber\\
-\frac{\alpha}{M}{\bf m}\times\frac{\partial{\bf m}}{\partial t}.
\label{LLG}
\end{eqnarray}
The first term of the right-hand side (RHS) of (\ref{LLG}) relates to the exchange spin interaction while the second term
 depends on the external magnetic field ${\bf H}$; thus, $\gamma$ denotes the giromagnetic factor (up to its sign). 
 The constant $\beta_{1(2)}$ determines the strength of the easy axis (plane) anisotropy and
 $\hat{i}\equiv(1,0,0)$, $\hat{j}\equiv(0,1,0)$. Note that the long axis of the magnetic 
 nanowire is an easy axis for the majority of real systems; however, another choice of the anisotropy axes 
 does not influence the magnetization dynamics. The constant $\delta$ is proportional to the intensity 
 of the electric current through the wire and $\delta$ changes its sign under time-arrow reversal 
 (the inversion of the electron flow) \cite{slo96,baz98}. The non-adiabatic part of the current-induced 
 torque (which depends on $\beta$) is of dissipative origin; thus, $\beta\to 0$ with decreasing
 Gilbert damping constant $\alpha\to 0$ (one takes $\alpha,\beta\ll 1$). Its inclusion is necessary 
 if one describes an observed monotonous motion of the DW below the Walker breakdown \cite{zha04}-\cite{bea08}.
 Notice that including the magnetic-dissipation term following the original LL approach 
 (changing the last term of (\ref{LLG}) into $-\alpha{\bf m}\times[{\bf m}\times{\bf h}_{\rm eff}]$, where
 $h_{{\rm eff}j}=-\delta{\cal H}/\delta m_{j}$, and ${\cal H}$ denotes the Hamiltonian) would lead 
 to changing the constant $\beta$ into $\beta-\alpha$, \cite{tse08}. Although the discussion 
 of the relevance of both approaches to the magnetic dissipation remains open \cite{tse08,sti07},
 I believe, the clinching argument for the Gilbert approach is the expectation for the proper
 dissipative term to be dependent on the time derivative of the dynamical parameter ${\bf m}$.
 In other case, the dissipative term could influence static solutions to the LL equation
 while one expects the magnetic friction to be the kinetic.

Since (\ref{LLG}) is valid only when the constraint $|{\bf m}|=M$ is satisfied, 
 I intend to write equations of the unconstrained dynamics equivalent to (\ref{LLG}). 
 Introducing the complex dynamical parameters $m_{\pm}=m_{y}\pm{\rm i}m_{z}$, I represent 
 the magnetization components using a pair of complex functions $g(x,t)$, $f(x,t)$. This way I
 reduce the number of independent degrees of freedom. The relation between 
 the primary and secondary dynamical variables  
\begin{eqnarray}
m_{+}=\frac{2M}{f^{*}/g+g^{*}/f},
\hspace*{2em}
m_{x}=M\frac{f^{*}/g-g^{*}/f}{f^{*}/g+g^{*}/f}
\label{transform}
\end{eqnarray}
[where $(\cdot)^{*}$ denotes the complex conjugate (c.c.)] ensures that $|{\bf m}|=M$ while there are no constraints 
 on $g$, $f$. The transform (\ref{transform}) enables bilinearization ("trilinearization") of (\ref{LLG}) 
 following the Hirota method of solving nonlinear equations \cite{bog80,hir82}.
 In the particular case $H_{y}=H_{z}=0$, from (\ref{LLG}) and (\ref{transform}), we arrive at 
 the trilinear equations for $f$, $g$
\begin{eqnarray}
f\left[-{\rm i}D_{t}+JD_{x}^{2}+\delta(\beta-{\rm i})D_{x}+\alpha D_{t}\right]f^{*}\cdot g
\nonumber\\
+Jg^{*}D_{x}^{2}g\cdot g
-\left(\gamma H_{x}+\beta_{1}+\frac{\beta_{2}}{2}\right)|f|^{2}g
\nonumber\\
-\frac{\beta_{2}}{2}f^{*2}g^{*}=0,
\nonumber\\
g^{*}\left[-{\rm i}D_{t}-JD_{x}^{2}+\delta(\beta-{\rm i})D_{x}+\alpha D_{t}\right]f^{*}\cdot g
\nonumber\\
-JfD_{x}^{2}f^{*}\cdot f^{*}
+\left(-\gamma H_{x}+\beta_{1}+\frac{\beta_{2}}{2}\right)|g|^{2}f^{*}
\nonumber\\
+\frac{\beta_{2}}{2}g^{2}f=0,
\label{secondary-eq}
\end{eqnarray}
where $D_{t}$, $D_{x}$ denote Hirota operators of differentiation which are defined by 
\begin{eqnarray}
D_{t}^{m}D_{x}^{n}b(x,t)\cdot c(x,t)\equiv
(\partial/\partial t-\partial/\partial t^{'})^{m}
\nonumber\\
\times
(\partial/\partial x-\partial/\partial x^{'})^{n}b(x,t)c(x^{'},t^{'})|_{x=x^{'},t=t^{'}}.
\nonumber
\end{eqnarray}

The inclusion of the dissipation into the LLG equation is connected to breaking the symmetry
 with relevance to the time reversal. Therefore, neither (\ref{LLG}) nor (\ref{secondary-eq}) 
 can describe the magnetization evolution on the whole time axis.
 In particular, the application of an external magnetic field or current to the DW system (or creation of a domain 
 in the presence of an external field) initiates a nonequilibrium process of the DW motion. Such a motion
 cannot be present in the distant past ($t\to-\infty$) since a nonzero value of the dissipative function 
 [relevant to the Gilbert term in (\ref{LLG})] would indicate unlimited growth of the energy with $t\to-\infty$.
 Thus, {\it solitary-wave solutions to (\ref{LLG}) are relevant only in the limit of large positive values 
 of time \cite{zha04,tse08}. This fact makes impossible an exact analysis of the DW collisions 
 using} (\ref{secondary-eq}) and motivates extension of the dynamical system within a formalism
 applicable to the whole length of the time axis. 
 
We write modified equations of motion using a similar trick to the one proposed 
 by Bateman with application to the Lagrangian description of the damped harmonic oscillator \cite{bat31}. 
 It is connected to the concept by Lakshmanan and Nakamura of removing the dissipative term from the evolution
 equation of ferromagnets via multiplying the time variable by a complex constant \cite{lak84}, however,
 it demands an improvement in the spirit of Bateman's idea \cite{mag86}. The concept is to extend the dynamical 
 system doubling the number of freedom degrees and adding the equations which differ from the original ones
 by the sign of the dissipation constants. Resulting extended system is symmetric with relevance 
 to the time-arrow reversal, however, its solution consists of physical and virtual fields.
 Let us mention that different quantum dissipative formalisms (non-equilibrium Green functions, thermo-field 
 dynamics, rigged Hilbert space) are based on the Bateman's trick \cite{kel65}.
 
We extend the system of secondary dynamical equations (\ref{secondary-eq}) since it describes unconstrained 
 dynamics unlike the primary LLG equation (\ref{LLG}). We replace $g$, $g^{*}$, $f$, $f^{*}$
 in (\ref{secondary-eq}) with novel fields of the corresponding set $g_{1}$, $g_{2}^{*}$, $f_{2}$, $f_{1}^{*}$ 
 and of the set of their c.c.. For $\alpha,\beta=0$, in the absence of dissipation, 
 $g_{1}=g_{2}=g$, $f_{1}=f_{2}=f$. For the case $H_{y}=H_{z}=0$, the secondary dynamical equations 
 transform into 
\begin{widetext}
\begin{eqnarray}
f_{2}\left[-{\rm i}D_{t}+JD_{x}^{2}+\delta(\beta-{\rm i})D_{x}+\alpha D_{t}\right]f_{1}^{*}\cdot g_{1}
+Jg_{2}^{*}D_{x}^{2}g_{1}\cdot g_{1}
-\left(\gamma H_{x}+\beta_{1}+\frac{\beta_{2}}{2}\right)f_{2}f_{1}^{*}g_{1}
-\frac{\beta_{2}}{2}f_{2}^{*}f_{1}^{*}g_{1}^{*}=0,
\nonumber\\
g_{2}^{*}\left[-{\rm i}D_{t}-JD_{x}^{2}+\delta(\beta-{\rm i})D_{x}+\alpha D_{t}\right]f_{1}^{*}\cdot g_{1}
-Jf_{2}D_{x}^{2}f_{1}^{*}\cdot f_{1}^{*}
+\left(-\gamma H_{x}+\beta_{1}+\frac{\beta_{2}}{2}\right)g_{2}^{*}g_{1}f_{1}^{*}
+\frac{\beta_{2}}{2}g_{2}g_{1}f_{1}=0.
\label{secondary-eq-pm1}
\end{eqnarray}
Writing (\ref{secondary-eq-pm1}), we have replaced the last terms on the lhs of (\ref{secondary-eq}) 
 in a way to be linear in $g_{2}^{(*)}$, $f_{2}^{(*)}$, which ensures that they vanish (diverge) with
 time in the presence of $H_{x}\neq 0$ with similar damping (exploding) rates as all other terms of these equations, 
 (in particular, their damping does not modify the anisotropy).
 The additional equations of the dynamical system differ from (\ref{secondary-eq-pm1}) by 
 the sign of the dissipation constants $\alpha$, $\beta$ 
\begin{eqnarray}
f_{1}\left[-{\rm i}D_{t}+JD_{x}^{2}+\delta(-\beta-{\rm i})D_{x}-\alpha D_{t}\right]f_{2}^{*}\cdot g_{2}
+Jg_{1}^{*}D_{x}^{2}g_{2}\cdot g_{2}
-\left(\gamma H_{x}+\beta_{1}+\frac{\beta_{2}}{2}\right)f_{1}f_{2}^{*}g_{2}
-\frac{\beta_{2}}{2}f_{1}^{*}f_{2}^{*}g_{2}^{*}=0,
\nonumber\\
g_{1}^{*}\left[-{\rm i}D_{t}-JD_{x}^{2}+\delta(-\beta-{\rm i})D_{x}-\alpha D_{t}\right]f_{2}^{*}\cdot g_{2}
-Jf_{1}D_{x}^{2}f_{2}^{*}\cdot f_{2}^{*}
+\left(-\gamma H_{x}+\beta_{1}+\frac{\beta_{2}}{2}\right)g_{1}^{*}g_{2}f_{2}^{*}
+\frac{\beta_{2}}{2}g_{1}g_{2}f_{2}=0.
\label{secondary-eq-pm2}
\end{eqnarray}
\end{widetext}
Though the previous dynamical variables $g(f)$, $g^{*}(f^{*})$ were mutually independent, they had to be c.c. to each other
 in order that the system of equations (\ref{secondary-eq}) and their c.c. was closed. In the system of eight equations; 
 (\ref{secondary-eq-pm1})-(\ref{secondary-eq-pm2}) and their c.c., $g_{1}(f_{1})$ is not a c.c. to $g_{2}(f_{2})$, while
 comparing (\ref{secondary-eq-pm1}) and (\ref{secondary-eq-pm2}), one sees that $g_{2}(x,t)$ ($f_{2}(x,t)$) 
 can be obtained from $g_{1}(x,t)$ ($f_{1}(x,t)$) via changing the sign of its parameters $\alpha$, $\beta$.  
 
Under the time-arrow inversion, the system of the novel equations transforms into itself if one accompanies
 this operation by the transform of the novel dynamical variables $g_{1(2)}\to f_{2(1)}$, $f_{1(2)}\to-g_{2(1)}$.
 The equations (\ref{secondary-eq-pm1}) and their c.c., which determine the magnetization dynamics 
 for large positive values of time (in particular, for $t\to\infty$), contain the differentials of the functions 
 $g_{1}$, $g_{1}^{*}$, $f_{1}$, $f_{1}^{*}$. Therefore, the magnetization vector should be expressed 
 with these functions in the relevant time regime. Writing the magnetization in the form
\begin{eqnarray}
m_{+}=\frac{2M}{f_{1}^{*}/g_{1}+g_{1}^{*}/f_{1}},\hspace{1em}
m_{x}=M\frac{f_{1}^{*}/g_{1}-g_{1}^{*}/f_{1}}{f_{1}^{*}/g_{1}+g_{1}^{*}/f_{1}}
\label{distant-future}
\end{eqnarray} 
ensures that their components satisfy $|{\bf m}|=M$, $m_{x}=m_{x}^{*}$, and they reproduce (\ref{transform}) for $\alpha=\beta=0$.
 In the regime of large negative values of time, in particular, for $t\to-\infty$, we can analyze the evolution 
 of the magnetization with the inversed time arrow. It is described with the reversed magnetization vector
\begin{eqnarray}
\tilde{m}_{+}=-\frac{2M}{f_{2}^{*}/g_{2}+g_{2}^{*}/f_{2}},\hspace{0.5em}
\tilde{m}_{x}=-M\frac{f_{2}^{*}/g_{2}-g_{2}^{*}/f_{2}}{f_{2}^{*}/g_{2}+g_{2}^{*}/f_{2}}.
\label{distant-past}
\end{eqnarray}

\section{Domain-wall motion}

Let us analyze the multi-domain solutions to (\ref{secondary-eq-pm1})-(\ref{secondary-eq-pm2})
 in the absence of any external magnetic field, ${\bf H}=0$. We search for the solutions 
 which describe a single DW and two DWs in the forms
 $f_{1}^{*}=1$, $g_{1}=w_{1}{\rm e}^{k_{1}x-l_{1}t}$, and 
 $f_{1}^{*}=1+v^{*}{\rm e}^{k_{1}x-l_{1}t}{\rm e}^{k_{2}x-l_{2}t}$,
 $g_{1}=w_{1}{\rm e}^{k_{1}x-l_{1}t}+w_{2}{\rm e}^{k_{2}x-l_{2}t}$, respectively,
 where $k_{j}={\rm Re}(k_{j})$, and ${\rm sign}(k_{1})=-{\rm sign}(k_{2})$. Inserting these ansatz into 
 (\ref{secondary-eq-pm1})-(\ref{secondary-eq-pm2}), one finds 
\begin{eqnarray}
l_{j}=\frac{1}{1+{\rm i}\alpha}\left\{
-\sqrt{-\left(Jk_{j}^{2}-\beta_{1}\right)\left[Jk_{j}^{2}-(\beta_{1}+\beta_{2})\right]}
\right.\nonumber\\\biggl.
+\delta(1+{\rm i}\beta)k_{j}\biggr\},
\nonumber\\
k_{j}=\sqrt{\frac{\beta_{1}+\beta_{2}(w_{j}+w_{j}^{*})^{2}/(4|w_{j}|^{2})}{J}}.
\end{eqnarray}
The single-wall (two-domain) solutions with $|k_{1}|\in(\sqrt{\beta_{1}/J},\sqrt{(\beta_{1}+\beta_{2})/J})$
 describe moving solitary waves (topological solitons), \cite{bog80,kos90}. One sees the correspondence between
 the wall width and the spin deviation from the easy plain since the dependence of $k_{j}$ on $w_{j}$.
 When $\delta=0$, static two-domain solutions represent the Bloch DW,
 ($|k_{1}|=\sqrt{\beta_{1}/J}$, $w_{1}=-w_{1}^{*}$), or Neel DW, ($|k_{1}|=\sqrt{(\beta_{1}+\beta_{2})/J}$,
 $w_{1}=w_{1}^{*}$), respectively (the nomenclature of \cite{kos90} which differs from Neel and Bloch wall definition
 of e.g. \cite{mid63}). These two solutions correspond to the ones found in \cite{bul64,sar76} within the XY model. 
 The DW profiles can be described with the functions 
\begin{eqnarray} 
m_{+}(x,t)&=&M\frac{w_{1}{\rm e}^{-{\rm i}{\rm Im}l_{1}t}}{|w_{1}|}{\rm sech}[k_{1}x-{\rm Re}l_{1}t+{\rm log}|w_{1}|],
\nonumber\\
m_{x}(x,t)&=&-M{\rm tanh}[k_{1}x-{\rm Re}l_{1}t+{\rm log}|w_{1}|],
\label{profile1}
\end{eqnarray}
($k_{1}>0$ relates to the head-to-head structure while $k_{1}<0$ to the tail-to-tail structure). 
 At the time points of the discrete set $t=\pi n/{\rm Im}l_{1}$, $n=0,\pm 1,\pm 2,\ldots$, one finds
 $g_{1}=g_{2}=g$, $f_{1}=f_{2}=f$ and (\ref{secondary-eq-pm1}) coincide with (\ref{secondary-eq}).
 Therefore, (\ref{profile1}) is a solitary-wave solution to (\ref{LLG}), (in particular, it coincides 
 with the one representing a Bloch DW or a Neel Dw for $\delta\neq 0$, \cite{thi05}). 
 Throughout the paper, we focus our attention on the externally-driven dynamics of the Bloch and Neel DWs,
 since these initially-static structures are the most important with relevance to the magnetic data storage.
 
We establish that static double-wall (three-domain) solutions to the LLG equation cannot 
 be written with the above Hirota expansion when $k_{1}=-k_{2}$. In this case the coefficient $v$;
\begin{eqnarray} 
v=-\frac{\beta_{2}Jk_{1}^{2}w_{1}^{*}w_{2}^{*}}{(Jk_{1}^{2}-\beta_{1})(Jk_{1}^{2}-\beta_{1}-\beta_{2})}
\end{eqnarray} 
diverges with $|k_{1}|\to\sqrt{\beta_{1}/J}$ or $|k_{1}|\to\sqrt{(\beta_{1}+\beta_{2})/J}$. Analogously to
 the XY model, the Hirota expansion is inapplicable to static three-domain configurations
 of the Bloch walls or Neel walls, while there exists static solution 
 to (\ref{secondary-eq-pm1})-(\ref{secondary-eq-pm2}) which describes a pair of different-type 
 (Neel and Bloch) walls \cite{bar05}. In particular, for $k_{1}=\sqrt{\beta_{1}/J}$, $k_{2}=-\sqrt{(\beta_{1}+\beta_{2})/J}$, 
 and $w_{1}=-w_{1}^{*}$, $w_{2}=w_{2}^{*}$, one finds
\begin{eqnarray} 
v=-\frac{\beta_{2}w_{1}w_{2}}{2\beta_{1}+\beta_{2}-2\sqrt{\beta_{1}(\beta_{1}+\beta_{2})}}
\end{eqnarray}  
Let us emphasize that we have not excluded the coexistence of a pair of Neel walls or Bloch walls in a magnetic wire. 
 However, the overlap of both the topological solitons induces their interaction which leads to 
 an instability of their parameters and, unlike for nontopological solitons, is not a temporal one \cite{bar07}. 
 
Solving (\ref{secondary-eq-pm1})-(\ref{secondary-eq-pm2}) in the presence of a longitudinal magnetic field 
 $H_{x}\neq 0$, we apply the ansatz
\begin{eqnarray}
f_{1}^{*}&=&\left(1+v^{*}{\rm e}^{k_{1}x-l_{1}t}{\rm e}^{k_{2}x-l_{2}t}\right){\rm e}^{\gamma H_{x}t/(-2{\rm i}+2\alpha)},
\nonumber\\
g_{1}&=&\left(w_{1}{\rm e}^{k_{1}x-l_{1}t}+w_{2}{\rm e}^{k_{2}x-l_{2}t}\right)
{\rm e}^{-\gamma H_{x}t/(-2{\rm i}+2\alpha)}
\label{external-ansatz}
\end{eqnarray} 
at the discrete time points $t=t_{n}\equiv 4\pi n(1+\alpha^{2})/(\gamma H_{x})$, where $n=0,\pm1,\pm2\ldots$,
 (let $\delta=0$ for simplicity). Behind these time points, in the presence of the 
 longitudinal field, the last terms on the lhs of (\ref{secondary-eq-pm1})-(\ref{secondary-eq-pm2}) change faster 
 (they oscillate with the three-times higher frequency) than the other ones. Therefore, taking the above ansatz,
 we apply an approach similar to the 'rotating wave approximation' in the quantum optics. Since this ansatz
 describes the spin structure rotation about the $x$-axis, it is applicable when the external field exceeds 
 Walker-breakdown critical value, $|H_{x}|>H_{W}$. From the single-wall solution, (the case of $w_{2}=0$ 
 or $w_{1}=0$), for $k_{1}=\sqrt{\beta_{1}/J}$, $k_{2}=-\sqrt{(\beta_{1}+\beta_{2})/J}$, we establish that
 applying the magnetic field in the easy-axis direction drives the DW motion with the velocity 
\begin{eqnarray}
c_{1(2)}=\gamma|H_{x}|\alpha/[|k_{1(2)}|(1+\alpha^{2})].
\label{velocity2}
\end{eqnarray}
Correspondingly, for $H_{x}=0$, applying the electric current through the initially static wall 
 drives it to move with the velocity 
\begin{eqnarray}
c=\frac{\delta(1+\alpha\beta)}{1+\alpha^{2}}
\label{velocity}
\end{eqnarray} 
which is independent of the DW width. 

The essential difference between both kinds of the driven motion 
 emerges from the analysis of the three-domain solutions. Under the external field, the two consecutive DWs
 move in the opposite directions. The walls which are closing up to each other collide and eventually 
 they can annihilate or wander off each other. The application of the electric current along the magnetic wire
 drives both the DWs to move in the same direction with the same velocity.
 Analyzing long-time limits of the magnetization vector in different regions of the coordinate $x$, 
 we establish the consequences of the field-induced collision of the complex of a Bloch DW interacting with
 a Neel DW. We use the ansatz (\ref{external-ansatz}) and assume $\delta=0$. 
 
Let $\eta_{j}\equiv k_{j}(x-x_{0j})-\gamma H_{x}\alpha t/(1+\alpha^{2})$,
 $\tilde{\eta}_{j}\equiv k_{j}(x-x_{0j})+\gamma H_{x}\alpha t/(1+\alpha^{2})$. 
 For $H_{x}>0$, at $t=t_{n}$ (within the above 'rotating wave approximation'), 
 we find the distant-future limit of the magnetization (\ref{distant-future})
\begin{eqnarray}
m_{+}\approx\left\{\begin{array}{cc}
m_{+}^{(1)} & \eta_{2}\ll\eta_{1}\sim 0\\
m_{+}^{(2)} & \eta_{1}\ll\eta_{2}\sim 0
\end{array}\right.=\lim_{t\to\infty}m_{+},
\nonumber\\
m_{x}\approx\left\{\begin{array}{cc}
m_{x}^{(1)} & \eta_{2}\ll\eta_{1}\sim 0\\
m_{x}^{(2)} & \eta_{1}\ll\eta_{2}\sim 0
\end{array}\right.=\lim_{t\to\infty}m_{x},
\label{collision_1}
\end{eqnarray}
where
\begin{eqnarray}
m_{+}^{(j)}=2M\frac{v/w_{j}^{*}{\rm e}^{\tilde{\eta}_{k}}{\rm e}^{-{\rm i}\gamma H_{x}t/(1+\alpha^{2})}}{
1+|v|^{2}/|w_{j}|^2{\rm e}^{2\tilde{\eta}_{k}}},
\nonumber\\
m_{x}^{(j)}=-M\frac{1-|v|^{2}/|w_{j}|^2{\rm e}^{2\tilde{\eta}_{k}}}{1+|v|^{2}/|w_{j}|^2{\rm e}^{2\tilde{\eta}_{k}}},
\label{collision_1-prime}
\end{eqnarray}
and $j\neq k$. Identifying the parameters $x_{0j}$ with the DW-center positions, we introduce 
 the restriction on $w_{j}$, $|v|/|w_{j}|=1$. We notice that $m_{+}^{(1)}$, $m_{x}^{(1)}$ as well as 
 $m_{+}^{(2)}$, $m_{x}^{(2)}$ are the Walker single-DW solutions to the primary LLG equation which describe the motion
 of well separated DWs \cite{sch74,bea05}. Thus, our three-domain profiles of the fields 
 (\ref{distant-future}) tend to satisfy (\ref{LLG}) in the limit $t\to\infty$ according to the requirement
 formulated in the previous section. 
 
In the distant-past limit, we describe the magnetization evolution with the reversed time arrow. Following (\ref{distant-past}), 
\begin{eqnarray}
\tilde{m}_{+}\approx\left\{\begin{array}{cc}
\tilde{m}_{+}^{(1)} & \tilde{\eta}_{1}\ll\tilde{\eta}_{2}\sim 0\\
\tilde{m}_{+}^{(2)} & \tilde{\eta}_{2}\ll\tilde{\eta}_{1}\sim 0
\end{array}\right.=\lim_{t\to-\infty}\tilde{m}_{+},
\nonumber\\
\tilde{m}_{x}\approx\left\{\begin{array}{cc}
\tilde{m}_{x}^{(1)} & \tilde{\eta}_{1}\ll\tilde{\eta}_{2}\sim 0\\
\tilde{m}_{x}^{(2)} & \tilde{\eta}_{2}\ll\tilde{\eta}_{1}\sim 0
\end{array}\right.=\lim_{t\to-\infty}\tilde{m}_{x},
\end{eqnarray}
where
\begin{eqnarray}
\tilde{m}_{+}^{(j)}=-2M\frac{v/w_{k}^{*}{\rm e}^{\eta_{j}}{\rm e}^{-{\rm i}\gamma H_{x}t/(1+\alpha^{2})}}{
1+|v|^{2}/|w_{k}|^2{\rm e}^{2\eta_{j}}},
\nonumber\\
\tilde{m}_{x}^{(j)}=M\frac{1-|v|^{2}/|w_{k}|^2{\rm e}^{2\eta_{j}}}{1+|v|^{2}/|w_{k}|^2{\rm e}^{2\eta_{j}}},
\label{collision_2-prime}
\end{eqnarray}
and $j\neq k$. In order to consider the collision of the pair of DWs which are infinitely distant from each other 
 at the beginning of their evolution, we determine the magnetization dynamics in the limit $t\to-\infty$. 
 For this aim, one has to invert the propagation direction of the kinks of $\tilde{\bf m}$ and to reverse 
 the arrow's head of the field vector $\tilde{\bf m}$. Utilizing the properties 
 $\tilde{m}_{+}^{(j)}(x+x_{0k},0)=\tilde{m}_{+}^{(j)}(-x+x_{0k},0)$, 
 $\tilde{m}_{x}^{(j)}(x+x_{0k},0)=-\tilde{m}_{x}^{(j)}(-x+x_{0k},0)$, we arrive at   
\begin{eqnarray}
m_{+}(x,t)&=&\left\{\begin{array}{cc}
-\tilde{m}_{+}^{(1)}(-x+2x_{01},t) & \eta_{1}\gg\eta_{2}\sim 0\\
-\tilde{m}_{+}^{(2)}(-x+2x_{02},t) & \eta_{2}\gg\eta_{1}\sim 0
\end{array}\right.,
\nonumber\\
m_{x}(x,t)&=&\left\{\begin{array}{cc}
-\tilde{m}_{x}^{(1)}(-x+2x_{01},t) & \eta_{1}\gg\eta_{2}\sim 0\\
-\tilde{m}_{x}^{(2)}(-x+2x_{02},t) & \eta_{2}\gg\eta_{1}\sim 0
\end{array}\right..
\label{collision_2}
\end{eqnarray}
The applicability of the above procedure to study the asymptotic evolution of 
 a single DW is easy to verify since any single-DW solution satisfies
\begin{eqnarray}
m_{+}(x,t)=-\tilde{m}_{+}(-x+2x_{01},t),
\nonumber\\
m_{x}(x,t)=\tilde{m}_{x}(-x+2x_{01},t).
\end{eqnarray}
Typically, one should consider the formulas (\ref{collision_1}), (\ref{collision_2})
 with relevance to the case $\beta_{1}\gg\beta_{2}$, thus $k_{1}\approx-k_{2}$, which 
 corresponds to commonly studied crystalline magnetic nanowires, e.g. for Fe, FePt, 
 $\beta_{2}/\beta_{1}\sim10^{-1}$, \cite{ved04}. For noncrystalline (permalloy) nanowires 
 Fe$_{1-x}$Ni$_{x}$ deposited on a crystalline substrate, the easy-axis anisotropy constant
 determined from uniform-resonance measurements was found to be, unexpectedly, as big as 
 in the crystalline nanowires \cite{alo05}. Therefore, even when neglect structural effects 
 in real systems which lead to the saturation of the spin alignment in the DW area to the easy plain 
 or hard plain (the Walker breakdown) which suppresses their spontaneous motion,
 the spectrum of spontaneously propagating DWs in nanowires would be
 very narrow and their velocities would be very small. 

According to (\ref{collision_1}), (\ref{collision_2}), two initially closing up DWs have to diverge after 
 the collision. If one of the colliding DWs that was initially,
 for $t\to-\infty$, described with the field ingredient $\tilde{m}_{+}^{(j)}$, $\tilde{m}_{x}^{(j)}$, it is finally, for $t\to\infty$, 
 described with the field ingredient $m_{+}^{(j)}$, $m_{x}^{(j)}$. 
 Therefore, reflecting
 DWs exchange their parameters $x_{01}\leftrightarrow x_{02}$, $w_{1}\leftrightarrow w_{2}$, $k_{1}\leftrightarrow-k_{2}$.
 It is connected to exchanging the directions of the spin orientation in the $yz$-plane in the wall areas, (after 
 the collision the Neel wall changes into the Bloch wall and vice versa as shown in Fig. 1). 
 Our prediction corresponds to the result 
 of the collision analysis performed for spontaneously propagating DWs (in absence of external field, 
 electric current, and dissipation). According to findings of \cite{kos90,liu02}, the DWs reflect during 
 the collision in a way that one can say they pass through each other without changing their widths
 and velocities, however, with changing their character from the head-to-head one into the tail-to-tail 
 one and vice versa. 

\begin{figure}
\unitlength 1mm
\begin{center}
\begin{picture}(175,40)
\put(0,-5){\resizebox{85mm}{!}{\includegraphics{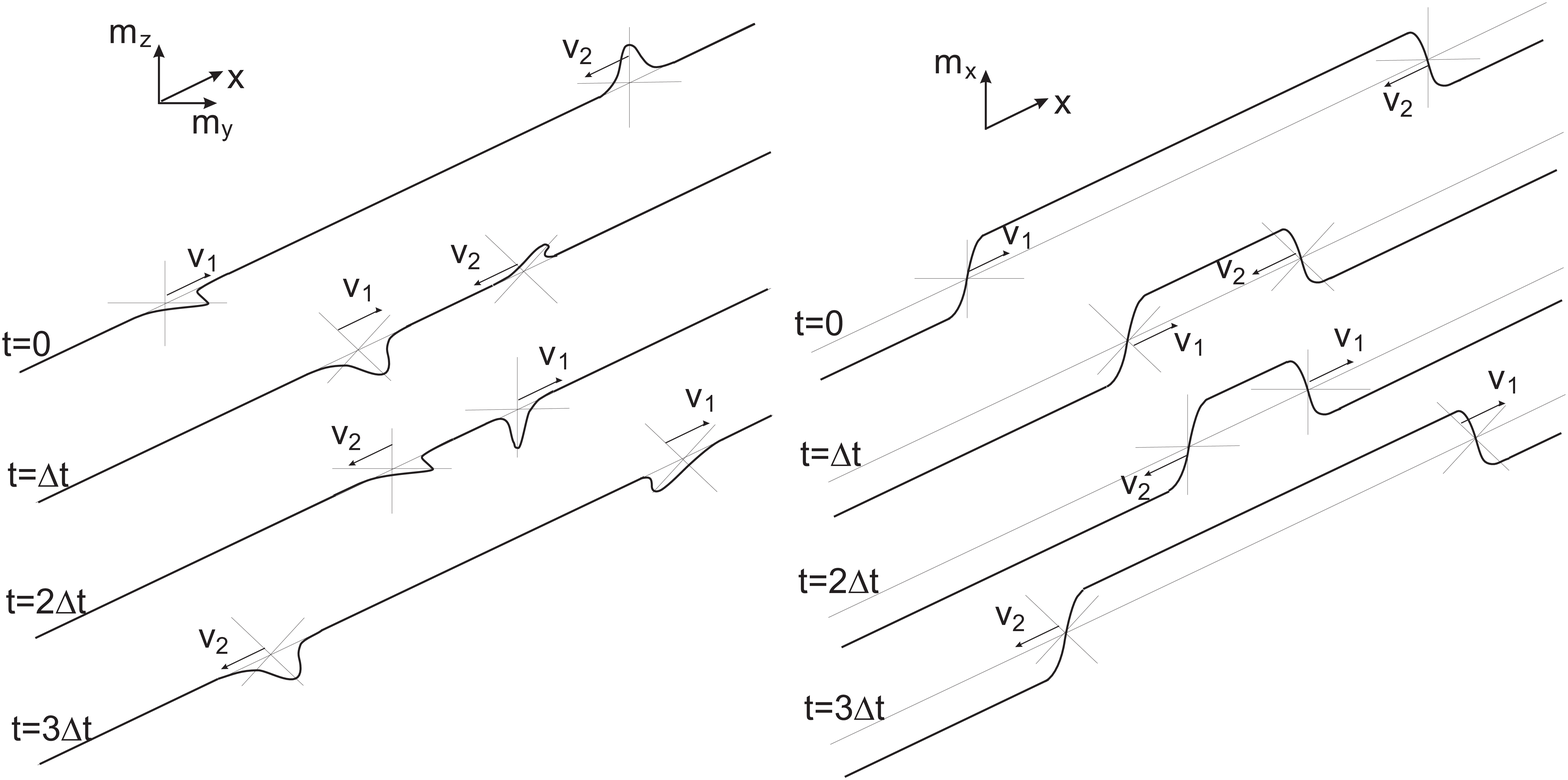}}}
\end{picture}
\end{center}
\caption{The magnetization dynamics of a system of one Neel DW and one Bloch DW in a longitudinal field above
the Walker breakdown. Their reflection takes place in the time region $(\Delta t,2\Delta t)$ and
it is accompanied by a change of the Bloch wall (of the velocity $v_{2}$) into the Neel wall (of the velocity $v_{1}$) 
and vice versa. Since $|H_{x}|>H_{W}$, the spin structure monotonously rotates about the $x$-axis.}
\end{figure}

Up till now, we have considered systems of infinite domains whose energy cannot be defined. However, the smaller 
 a domain is the bigger percentage of the Zeeman part of its energy is lost per time unit due to the DW motion. 
 The condition of the domain-energy minimization determines the direction of this motion. The domains aligned 
 parallelly to the external field grow while the domains aligned antiparallelly to the field diminish. 
 Any DW reflection induces a motion which contradicts this rule. Such a motion has to be decelerated and, eventually,
 it has to be suppressed when the decrease of the DW interaction energy equals the increase of the Zeeman energy. 
 The outcome of a many-collision process in a finite-size system is the appearance of a 1D magnetic-bubble structure 
 similar to widely known 2D bubble structures \cite{lee80}. The bubble size and concentration depend on the magnetic
 field intensity. Each bubble is ended with a Neel DW at one of its sides and with a Bloch DW at the other side.
 Some analogy with a complex boundary of hard (quasi-2D) magnetic bubbles can be noticed since such a border contains
 alternating Neel and Bloch points in its structure \cite{ros72}. Let us mention, that interesting concepts of storing
 and transforming the binary information have had been developed several decades ago with relevance to 2D magnetic 
 bubble systems, though, they have been abandoned because of technological problems of the time \cite{ode86}.   
 
Numerical analyses of the field-induced DW collision (micromagnetic studies) have been performed 
 with relevance to flattened nanowires (quasi 1D nanostripes) below and just above the Walker
 breakdown using the dissipative LL equation \cite{kun09}. 
 The systems below the threshold correspond to a plane-rotator model studied in section 4, while the systems 
 above the threshold are qualitatively described with the present model. The mentioned simulations focus
 on the collisions of similar-type (Neel or Bloch) DWs neglecting the anisotropy. They have predicted
 mutual annihilation or reflection of the walls depending on (parallel or anti-parallel) spin alignments
 in both the DW centers. This result is partially supported by a perturbation analysis of
 Bloch-wall interactions within the XY model which has shown such DWs to repel or attract each other
 depending on their chiralities \cite{bar07}. The method of present study cannot be applied to these collisions
 since one is unable to determine neither the double-Neel nor double-Bloch wall analytical solutions 
 to the dynamical equations. However, except in the case of a periodically distributed DWs, 
 multi-Bloch or multi-Neel structures are unstable because of unbalanced DW interactions, thus, they seem
 to be less suitable for the information-storing purposes than the Neel-Bloch DW structures.
 To the best of the author knowledge, the field-induced collision of the Neel DW with the Bloch DW
 has not been simulated. 
 
\section{Plane-rotator model}

In order to describe the DW dynamics below the Walker breakdown, we consider a system of plane rotators. 
 Let us reduce the primary (LLG) dynamical system to its single component. Saturating the magnetization dynamics 
 to the easy plain ($m_{y}=0$), we neglect the spin rotation about the $x$-axis and $z$-axis since
 the relevant torque components are equal to zero. For ${\bf H}=(H_{x},0,0)$, inserting 
\begin{eqnarray}
m_{x}=M\frac{1-a_{1}^{2}}{1+a_{1}^{2}},\hspace*{1em}m_{y}=0,\hspace*{1em}m_{z}=2M\frac{a_{1}}{1+a_{1}^{2}}
\label{planar-ansatz}
\end{eqnarray}
(where $a_{1}$ takes real values) into the $y$-component of (\ref{LLG}), one arrives at a nonlinear diffusion
 equation
\begin{eqnarray}
\left(-\alpha\frac{\partial a_{1}}{\partial t}-\gamma H_{x}a_{1}
-\delta\beta\frac{\partial a_{1}}{\partial x}+J\frac{\partial^{2}a_{1}}{\partial x^{2}}\right)
\left(1+a_{1}^{2}\right)
\nonumber\\
-2Ja_{1}\left(\frac{\partial a_{1}}{\partial x}\right)^{2}
-\beta_{1}a_{1}\left(1-a_{1}^{2}\right)=0.
\label{diffusion}
\end{eqnarray}
We use another ansatz describing the dynamics constrained to the $xy$-plain (a hard plain)
\begin{eqnarray}
m_{x}=M\frac{1-a_{2}^{2}}{1+a_{2}^{2}},\hspace*{1em}m_{y}=2M\frac{a_{2}}{1+a_{2}^{2}},\hspace*{1em}m_{z}=0.
\label{planar-ansatz'}
\end{eqnarray}
Then, we insert it into the $z$-component of (\ref{LLG}) and we arrive at
\begin{eqnarray}
\left(-\alpha\frac{\partial a_{2}}{\partial t}-\gamma H_{x}a_{2}
-\delta\beta\frac{\partial a_{2}}{\partial x}+J\frac{\partial^{2}a_{2}}{\partial x^{2}}\right)
\left(1+a_{2}^{2}\right)
\nonumber\\
-2Ja_{2}\left(\frac{\partial a_{2}}{\partial x}\right)^{2}
-(\beta_{1}+\beta_{2})a_{2}\left(1-a_{2}^{2}\right)=0
\label{diffusion'}
\end{eqnarray}
which differs from (\ref{diffusion}) by a constant at the anisotropy term.
 With relevance to the case $\delta=0$, one finds the two-domain solution
\begin{eqnarray}
a_{1(2)}=w{\rm e}^{k_{1(2)}x-\gamma H_{x}t/\alpha},
\nonumber\\
|k_{1}|=\sqrt{\frac{\beta_{1}}{J}},
\hspace*{1em}|k_{2}|=\sqrt{\frac{\beta_{1}+\beta_{2}}{J}},
\end{eqnarray}
which correspond to the Bloch DW (to the Neel DW). When $H_{x}\neq 0$, the DW propagates with the velocity
\begin{eqnarray} 
c_{1(2)}=\frac{\gamma|H_{x}|}{|k_{1(2)}|\alpha}.
\label{velocity1}
\end{eqnarray}

The applicability of the plane-rotator model is limited by the Walker-breakdown condition. The magnetic 
 field $|H_{x}|$ cannot exceed a critical value $H_{W}$ (see Fig. 2a) which corresponds 
 to the spin deviation from the basic magnetization plane (a canting) at the center of the DW about 
 a limit angle equal or smaller than $\pi/4$. We estimate an upper limit of the Walker critical field 
 considering the $x$-component of the LLG equation at the DW center 
\begin{eqnarray}
\frac{\partial m_{x}}{\partial t}\bigg|_{x=x_{01(2)}}\approx\left[\frac{\beta_{2}}{M}m_{y}m_{z}
-\delta\frac{\partial m_{x}}{\partial x}\right]\bigg|_{x=x_{01(2)}},
\label{center_canting}
\end{eqnarray}
where $x_{01(2)}\equiv-\log(w)/k_{1(2)}$. Let $\varphi$ denotes the angle of the spin deviation (a canting) at 
 DW center. Inserting (\ref{planar-ansatz}) and transforming $m_{y}m_{z}\to m_{z}^{2}\sin(\varphi)\cos(\varphi)$ in 
 (\ref{center_canting}) or inserting (\ref{planar-ansatz'}) and transforming
 $m_{z}m_{y}\to m_{y}^{2}\sin(\varphi)\cos(\varphi)$ in (\ref{center_canting}), one arrives at
\begin{eqnarray}
\frac{\partial a_{1(2)}}{\partial t}=\frac{\beta_{2}}{2}\sin(2\varphi)a_{1(2)}-\delta\frac{\partial a_{1(2)}}{\partial x}
\label{Walker}
\end{eqnarray}
and finally, assuming $|\varphi|\le\pi/4$, at 
\begin{eqnarray}
|H_{x}|\le H_{W}\le{\rm max}H_{W}\equiv\frac{\alpha\beta_{2}}{2\gamma}+\frac{\alpha\delta}{\gamma}|k_{2}|.
\end{eqnarray}
This expression corresponds to the one given in \cite{hay06,bea05,mou07}. However, we notice that, for typical nanowires
 whose width-to-thickness ratio is bigger than 20 (double-atomic or triple-atomic layers of a submicrometer width), 
 measuring $H_{W}$, one has estimated the canting angle to take a value of a few degrees 
 at most, \cite{bea08,bea05,kun06}. 

For ${\bf H}=0$, $\delta\neq 0$, the two-domain solution to (\ref{diffusion})-(\ref{diffusion'}) takes the form 
\begin{eqnarray}
a_{1(2)}=w{\rm e}^{k_{1(2)}(x-\delta\beta t/\alpha)},
\nonumber\\|k_{1}|=\sqrt{\frac{\beta_{1}}{J}},
\hspace*{1em}|k_{2}|=\sqrt{\frac{\beta_{1}+\beta_{2}}{J}}.
\end{eqnarray}
It is seen that only the non-adiabatic part of the spin-transfer torque contributes 
 to (\ref{diffusion}),(\ref{diffusion'}) since the current-dependent term is proportional to $\beta$.
 From (\ref{Walker}), the current-induced Walker breakdown corresponds to the critical current intensity 
\begin{eqnarray}
\delta_{W}\le\beta_{2}/[2|k_{1}|(1+\beta/\alpha)]
\label{critical-current}
\end{eqnarray}
 if $H_{x}=0$, (see Fig. 2b). 
 It has been observed that $H_{W}$, $\delta_{W}$ decrease with decreasing the nanowire width-to-thickness 
 ratio, \cite{kun06,gla08}, because this ratio determines the strength of the easy-plain anisotropy while
 $H_{W},\delta_{W}\to 0$ with $\beta_{2}\to 0$, \cite{yan10}. 
 Notice that analytical calculations using 2D XY model,
 experimental observations, and simulations of the spin ordering in nanostripes
 show this ordering to vary along the cross-section width of the nanostripe in the DW area, thus,
 revealing a complex topological structure \cite{kun09,yua91}. Therefore, our plane-rotator
 description is valid only for a qualitative analysis of the DW dynamics in the nanostripes. 

\begin{figure}
\unitlength 1mm
\begin{center}
\begin{picture}(160,35)
\put(0,-5){\resizebox{85mm}{!}{\includegraphics{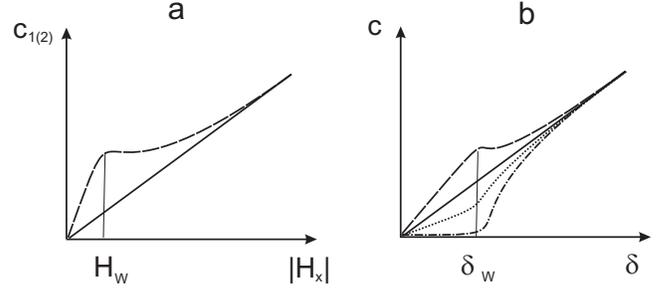}}}
\end{picture}
\end{center}
\caption{a) A scheme of the longitudinal-field dependence of the DW velocity for: a wire with single-axis
 anisotropy (solid line), a wire with double-axis anisotropy (dashed line). b) A scheme of the current-intensity dependence
of the DW velocity for: a wire with single-axis anisotropy (solid line), a wire with double-axis anisotropy; 
 $\beta>\alpha$ (dashed line), $\beta<\alpha$ (dotted line), $\beta=0$ (dash-dotted line).}
\end{figure}
 
Neither finding nonstationary double-Bloch nor double-Neel solutions in the form of the Hirota expansion
 (including its second order) does not manage. In particular, inserting 
\begin{eqnarray}
a_{1(2)}=\frac{w_{1}{\rm e}^{k_{1(2)}x}+w_{2}{\rm e}^{k_{1(2)}^{'}x}}{1+v_{1(2)}{\rm e}^{k_{1(2)}x+k_{1(2)}^{'}x}}
{\rm e}^{-\gamma H_{x}t/\alpha},
\end{eqnarray}
into (\ref{diffusion})-(\ref{diffusion'}), for $k_{1}=-k_{1}^{'}=\pm\sqrt{\beta_{1}/J}$,
 $k_{2}=-k_{2}^{'}=\pm\sqrt{(\beta_{1}+\beta_{2})/J}$, leads to the divergence of $v_{1(2)}$ as it 
 follows from the approach of section 3. In order to describe the collision of Bloch and Neel walls below the 
 Walker breakdown, I propose to apply an effective 1D model assuming the magnetization precession to be overdamped, 
 thus, taking the lhs of (\ref{LLG}) to be equal to zero. Inclusion of the constraint $|{\bf m}|=M$ leads to 
 the modified (by neglecting the first terms on the lhs) system (\ref{secondary-eq-pm1})-(\ref{secondary-eq-pm2}).
 Solving it, we predict the field-induced DW collision 
 below the Walker breakdown to result in their reflection similar to the one described in section 3.
 The reflection is accompanied by the change of the Bloch wall into the Neel wall and vice versa.
 Let us emphasize that there is no spontaneous DW motion below the Walker breakdown when neglect
 magnetostatic effects \cite{dju10}. 

The technological challenge of increasing the DW speed is especially important below the Walker breakdown,
 where the driving-field is relatively weak. Referring to this purpose, we mention 
 an attempt utilizing an increase of the nanostripe-edge roughness, thus, an increase of the damping 
 constant $\alpha$, \cite{nak03}. This approach fails since, according to simulations of \cite{kun06},
 the maximum of the field-induced DW velocity is insensitive to $\alpha$ below the breakdown. It is because 
 $c_{1(2)}\propto\alpha^{-1}|H_{x}|$ while $|H_{x}|\le H_{W}\propto\alpha$. On the other hand, since $\beta$ 
 grows with $\alpha$, (the non-adiabatic part of the spin-transfer torque is of a dissipative origin), 
 the velocity of the current-induced DW motion 
\begin{eqnarray}
c=\delta\beta/\alpha
\label{velocity'}
\end{eqnarray}
 can be insensitive to the increase of the nanostripe-edge roughness as well.
 One has attributed some reported DW-velocity increase due to the nanostripe-edge roughness 
 to a decrease of its effective cross-section width. Another attempt utilized an increase of $H_{W}$ due to 
 the increase of the anisotropy constant $\beta_{2}$. It has been done via the nanowire deposition
 on a specific crystalline substrate \cite{lee07a}. However, the most efficient method of influencing 
 the maximum DW velocity below the Walker breakdown is the application of the transverse magnetic field,
 which is considered in the next section \cite{kun08}. 
 
\section{Domain wall in perpendicular to easy axis field}

Let us define $H_{\pm}\equiv H_{y}\pm{\rm i}H_{z}$. For $H_{x}=0$, we search for a two-domain solution 
 to (\ref{LLG}) using a different 'multi-linearization' than used in the previous sections 
\begin{eqnarray}
f_{2}\left[-{\rm i}D_{t}+JD_{x}^{2}+\delta(\beta-{\rm i})D_{x}+\alpha D_{t}\right]f_{1}^{*}\cdot g_{1}
\nonumber\\
+\frac{\gamma H_{+}}{2}f_{1}^{*}\left(f_{1}^{*}f_{2}+g_{2}^{*}g_{1}\right)
-\left(\beta_{1}+\frac{\beta_{2}}{2}\right)f_{2}f_{1}^{*}g_{1}
\nonumber\\
-\frac{\beta_{2}}{2}f_{1}^{*2}g_{2}^{*}&=&0,
\nonumber\\
g_{2}^{*}\left[-{\rm i}D_{t}-JD_{x}^{2}+\delta(\beta-{\rm i})D_{x}+\alpha D_{t}\right]f_{1}^{*}\cdot g_{1}
\nonumber\\
-\frac{\gamma H_{-}}{2}g_{1}\left(f_{1}^{*}f_{2}+g_{2}^{*}g_{1}\right)
+\left(\beta_{1}+\frac{\beta_{2}}{2}\right)g_{2}^{*}g_{1}f_{1}^{*}
\nonumber\\
+\frac{\beta_{2}}{2}g_{1}^{2}f_{2}&=&0,
\nonumber\\
f_{2}g_{1}D_{x}^{2}f_{1}^{*}\cdot f_{1}^{*}-f_{1}^{*}g_{2}^{*}D_{x}^{2}g_{1}\cdot g_{1}&=&0.
\label{secondary-eq-hpm}
\end{eqnarray}
Let us focus our attention on the case $\alpha,\beta,\delta=0$ for simplity. Then one has $f_{1}=f_{2}=f$, 
 $g_{1}=g_{2}=g$, while in the general case the relations (\ref{distant-future}), (\ref{distant-past}) apply.
 We analyze the two cases of the external-field direction; the one parallel 
 to the easy plane $H_{+}={\rm i}H_{z}$, and the one perpendicular to the easy plane $H_{+}=H_{y}$.

In the case of $H_{+}={\rm i}H_{z}$, we apply the ansatz 
\begin{eqnarray}
f_{1}^{*}=f_{2}=q_{1}+s_{1}{\rm e}^{k_{1}x-l_{1}t},
\nonumber\\
g_{1}=-g_{2}^{*}={\rm i}\left(s_{1}+q_{1}{\rm e}^{k_{1}x-l_{1}t}\right), 
\label{solution-hpm}
\end{eqnarray}
where $k_{1}$, $q_{1}$, $s_{1}$ denote real constants, (the parameter $l_{1}$ can take complex values when 
 $\alpha\neq 0$). The solution in the form (\ref{solution-hpm}) describes the wall between two domains whose
 spins are deviated from the easy axis onto the external-field direction about an angle which grows with $|H_{+}|$.
 Inserting this ansatz into (\ref{secondary-eq-hpm}), one finds 
\begin{eqnarray}
s_{1}=\frac{\beta_{1}-\sqrt{\beta_{1}^{2}-\gamma^{2}H_{z}^{2}}}{\gamma H_{z}}q_{1}.
\end{eqnarray}
Considering the solutions which are static in the absence of the electric current, $l_{1}=0$ for $\delta=0$,
 one arrives at 
\begin{eqnarray}
|k_{1}|=\sqrt{\frac{\beta_{1}^{2}-\gamma^{2}H_{z}^{2}}{\beta_{1}J}}.
\label{perpendicular_width_1}
\end{eqnarray}

In the case of $H_{+}=H_{y}$, the ansatz relating to the deviation of the domain magnetization
 from the easy axis onto the external-field direction takes the form
\begin{eqnarray}
f_{1}^{*}=f_{2}=q_{2}+s_{2}{\rm e}^{k_{2}x-l_{2}t},
\nonumber\\
g_{1}=g_{2}^{*}=s_{2}+q_{2}{\rm e}^{k_{2}x-l_{2}t}. 
\end{eqnarray}
with real $k_{2}$, $q_{2}$, $s_{2}$. From (\ref{secondary-eq-hpm}), we find
\begin{eqnarray}
s_{2}=\frac{\beta_{1}+\beta_{2}-\sqrt{(\beta_{1}+\beta_{2})^{2}-\gamma^{2}H_{y}^{2}}}{\gamma H_{y}}q_{2}.
\end{eqnarray}
The static solutions correspond to
\begin{eqnarray}
|k_{2}|=\sqrt{\frac{(\beta_{1}+\beta_{2})^{2}-\gamma^{2}H_{y}^{2}}{(\beta_{1}+\beta_{2})J}}.
\label{perpendicular_width_2}
\end{eqnarray}

The transverse external field does not drive the DW motion even in the presence 
 of the magnetic dissipation ($\alpha\neq 0$). When the current through the wire and the dissipation
 are applied, under the transverse magnetic field, the DW moves with the velocity $c$ given by
 (\ref{velocity}), which is independent of the value of this field. Then the solution 
 to (\ref{secondary-eq-hpm}) satisfies the bilinearized LLG system (Eqs. (3) with additional 
 $H_{+}$-dependent terms) at the time points of the discrete set $t=\pi n/{\rm Im}l_{1}$,
 where $n=0,\pm 1,\pm 2,\ldots$, since $f_{1}=f_{2}=f$, $g_{1}=g_{2}=g$ at these points.
 Including an additional to $H_{+}$ longitudinal component of the magnetic field $H_{x}$ drives the DW motion. 
 For the realistic case $H_{W}\sim|H_{x}|<|H_{+}|\ll|\beta_{1}/\gamma|$, neglecting small contributions
 to the $H_{x}$-dependent part of the torque, one finds the velocity of such a DW propagation (\ref{velocity2}) 
 or (\ref{velocity1}) above and below the Walker breakdown, respectively, with $|k_{1(2)}|$ given 
 by (\ref{perpendicular_width_1}), (\ref{perpendicular_width_2}). This velocity nonlinearly 
 increases with $|H_{+}|$, \cite{gla08a}. Searching for $c_{1(2)}$, in the case $|H_{x}|<H_{W}$, additionally,
 I have taken the lhs of (\ref{LLG}) equal to zero as discussed in section 4. 
 The manipulation of $c_{1(2)}$ via the application of the transverse magnetic field is potentially useful 
 for speeding up the processing with a magnetically-encoded information. We also notice that the transverse-field 
 dependence of $|k_{1(2)}|$ enables influencing the magnitude of the critical current of the Walker breakdown 
 $\delta_{W}$, following (\ref{critical-current}).

\section{Conclusions}

We have analytically studied the DW dynamics in the presence of the external magnetic field 
 and the electric current along the magnetic wire within the LLG approach. 
 It has demanded overcoming the difficulty arising from breaking the time-reversal symmetry 
 by inclusion of the magnetic dissipation. We have removed this asymmetry of the dynamical system 
 by introducing additional (virtual) dynamical variables, which is a similar trick to the Lagrangian 
 approach to the damped harmonic oscillator. Determining a connection of the additional dynamical 
 variables to the evolution of the magnetization vector in specific ranges of time,
 we have analyzed the dynamics of a single DW and of a pair of DWs. 

The magnetic-field-induced velocities of the DWs, the formulas (\ref{velocity2}) 
 and (\ref{velocity1}), and the current-induced velocities (\ref{velocity}) and (\ref{velocity'})
 are found to correspond to the ones of the Walker approach above and below the breakdown, respectively.
 According to \cite{bar05}, static three-domain solutions to 1D LLG equation describe
 pairs of Neel and Bloch walls. For the purposes of the qualitative dynamics analysis of a number of DWs 
 below the Walker breakdown, especially of the Neel-Bloch pairs, we have proposed a dynamical equation which 
 differs from the LLG one by neglecting the lhs in (\ref{LLG}). Below and above the breakdown, 
 the neighboring Neel and Bloch walls move in the presence of the longitudinal external field 
 in the opposite directions. Their collision results in the DW reflection accompanied by the reorientation of 
 the Neel wall into the Bloch wall and vice versa. In other words, the DWs pass trough each other without 
 changing their widths and velocities, however, the head-to-head DW structure changes into the tail-to-tail 
 one and vice versa.
    
Our method is useful for the analysis of two-domain systems under the transverse (with respect to
 the easy axis) external field, which enables a verification of numerical 
 and experimental results \cite{kun08,gla08a,sob95}. A reorientation of the magnetic domains due to the
 transverse field induces a widening of the DW area up to the infinity when approach with the field intensity 
 to the coercivity value. The consequence of the transverse-field application is an increase of 
 the DW mobility (the ratio $c_{1(2)}/|H_{x}|$) and an increase of the critical current (a shift of the 
 current-driven Walker breakdown).  

\section*{Acknowledgements}

This work was partially supported by Polish Government Research Founds for 2010-2012 in the framework of Grant No. N N202 198039.

\end{document}